\documentclass[12pt]{article}
\usepackage[dvips]{graphics}

\newcommand{\bd}[1]{ \mbox{\boldmath $#1$} }
\begin{document}
\def\ii{\'\i}

\title{A schematic model for QCD.\\
 I: Low energy meson states.}

\author{
S. Lerma H.\thanks{e-mail: alerma@nuclecu.unam.mx},
S. Jesgarz \thanks{e-mail: jesgarz@nuclecu.unam.mx},
P. O. Hess$^1$ \thanks{e-mail: hess@nuclecu.unam.mx}, \\
O. Civitarese$^2$ \thanks{e-mail: civitare@fisica.unlp.edu.ar},
and M. Reboiro$^2$
\thanks{e-mail: reboiro@fisica.unlp.edu.ar},
\\
\\ {\small\it$^{1}$Instituto de Ciencias Nucleares, Universidad
Nacional Aut\'onoma de M\'exico,} \\
{\small\it Apdo. Postal 70-543, M\'exico 04510 D.F.} \\
{\small\it$^{2}$ Departamento de F\'{\i}sica, Universidad Nacional de La Plata, } \\
{\small\it c.c. 67 1900, La Plata, Argentina. } \\
$\;\;\;\;\;\;\;\;\;\;$\\ }
\maketitle

\noindent {\small {\bf Abstract}: A simple model for QCD is
presented, which is able to reproduce the meson spectrum at low
energy. The model is a Lipkin type model for quarks coupled to
gluons. The basic building blocks are pairs of quark-antiquarks
coupled to a definite flavor and spin. These pairs are coupled to
pairs of gluons with spin zero. The multiplicity problem, which
dictates that a given experimental state can be described in
various manners, is removed when a particle-mixing interaction is
turned on. In this first paper of a series we  concentrates on the
discussion of meson states at low energy, the so-called zero
temperature limit of the theory. The treatment of baryonic states
is indicated, also.
} \\


\section{Introduction}

QCD is the favored theory of the strong interactions. At low
energy, however, the description of the hadronic spectrum based on
QCD becomes difficult due to the non linear structure of the
theory. This non-perturbative regimen, in contrast to lattice
gauge calculations, can be explored by means of schematic models.
The use of such models is common to other fields of physics where
the many-body structure of the theory may be explored by
introducing effective degrees of freedom and their couplings.
Nuclear structure physics is one of these examples and it is as
complicated and involved as the low-energy domain of QCD. Like in
the case of nuclear structure physics, QCD descriptions based on
simple models may help in the understanding of basic concepts and
procedures. The Lipkin model \cite{lipkin} is one of the most
famous schematic models, and it helped substantially to appreciate
the importance of pairing two-body interactions as well as the
importance of collectivity in building the low-energy part of the
nuclear spectrum. An extended version of the Lipkin model was
applied to the description of pion condensates in nuclei
\cite{schutte}. A variety of many body techniques have been tested
with Lipkin-type models \cite{stuart,betabeta}. In \cite{stuart}
some realistic, less schematic, nuclear interactions, suitable to
describe various nuclear properties, were investigated in this
way. In Ref. \cite{pittel} a Lipkin model was applied to describe
a system of many quarks. As seen in these examples, the
predictions of schematic models can be also rather rich in their
complexities. This fact was shown, for a simple model of many
gluon systems, in Ref. \cite{gluons99}.

Until now, the only formalism which can handle QCD  from first
principles is the lattice gauge theory \cite{latticebook}.
Particularly, in many gluon systems, a good description is
obtained without considering finite volume effects \cite{lattice}.
The problem with lattice gauge calculations, to treat QCD at low
energies, is that only the lowest state, and in some cases also
the next to the lowest state, for a given spin, charge conjugation
and parity, can be calculated. Lattice theory is numerically quite
involved, and the inclusion of quarks and antiquarks brings in
additional problems which are difficult to solve. Effective models
of the hadrons, like the MIT model \cite{mit}, can help to shed
some light into the structure of QCD at low energy. In
\cite{gluons99} the spectrum of gluons, as obtained in
\cite{lattice}, was reproduced and the sequence of levels
explained by simple assumptions. In other works \cite{swan,bonn}
many body methods were used to describe the spectrum of QCD at low
energy. After these considerations, it is obvious that it would be
nice to have a model which: i) must be able to describe the basic
structure of QCD at low and high energy, and  ii) must be solbable
exactly. Probably, such a model does not exist, due to the
complicated structure of QCD. Nevertheless, one can try to
construct a model which comes as near as possible to QCD.

 The purpose of the present work is to
present a model which fulfills the above requirements. In Ref.
\cite{simple} the most simple version of such a model was
presented. Like the models mentioned at the beginning, it is based
on a Lipkin type model and it  consists of two levels for the
description of quarks and antiquarks. These quarks are coupled to
a boson level which describes gluon pairs coupled to spin zero.
The other gluon states are treated as spectators. The basic
ingredients of the model are the quark-antiquark pairs coupled to
flavor singlet and spin zero and gluon pairs with spin zero. In
Ref. \cite{simple} it is shown that the model is able to describe
the appearance of a quantum phase transition at zero temperature,
when the interaction is turned on, and a phase transition to the
non interacting case at non-vanishing temperature. In Ref.
\cite{simple} the basic features of the model were discussed. Only
flavor singlet and spin zero mesons were taken into account. The
appearance of a Goldstone boson was obtained for a sufficiently
strong interaction. This state consisted of a meson with negative
parity. Also, it possesses a very collective nature, i.e. it is a
superposition of  many particles (quarks, antiquarks and gluons)
states. The behavior of the model at high temperature was
discussed, together with some consequences for the Quark-Gluon
Plasma (QGP) in \cite{qgp,qmd}.

In the first part of the paper we will introduce the general form
of the model for the description of meson states. The discussion
will concentrate on the behavior of the model at low energy,
corresponding to the zero temperature regime of QCD. The study of
the high energy behavior and the transition to the Quark Gluon
Plasma \cite{qgp,qmd} will be presented in the forthcoming paper
of the series \cite{papII}. In section 2 the basic ingredients of
the model are introduced, with the proposition of a Hamilton
operator. Because of the difficulties to treat fermion pairs
exactly, we shall diagonalize them in a boson mapping scheme. The
basis used to deal with the bosonic images of the fermion pairs,
and the corresponding matrix elements of the proposed Hamiltonian,
are given in the same section 2.  There we show how to assign
charge conjugation and G-parity symmetries to the states belonging
to the basis. In section 3 the model is applied to the description
of the low energy meson spectrum. Conclusions are drawn in section
4.

\section{The Model}

As indicated in Ref. \cite{simple} the fermion sector is described
by a Lipkin type model \cite{lipkin}, consisting of two levels,
one at energy $-\omega_f$ and the other at energy $+\omega_f$ (see
Fig. 1). This is the Dirac picture for fermions, where antiquarks
are regarded as holes in the lower level. Alternatively, one also
describes quarks and antiquarks in the level at $+\omega_f$. The
quarks and antiquarks are coupled to a 1.6 GeV level, which is
occupied by gluon pairs with spin zero \cite{gluons99}, a value
which was obtained in the description of a many gluon system and
which was adjusted to lattice gauge calculations \cite{lattice}.
In consequence, we shall take the energy of the gluon pair state
as an externally fixed data. The value $\omega_f$ is fixed to one
third of the nucleon mass (0.33GeV). There are further gluon
states \cite{gluons99} which do not interact with the quarks and
antiquarks. These states will be treated as spectators and should
be taken into account in the final spectrum.

\begin{figure}[tph]

\rotatebox{270}{\resizebox{356pt}{356pt}{\includegraphics[100,130][612,660]{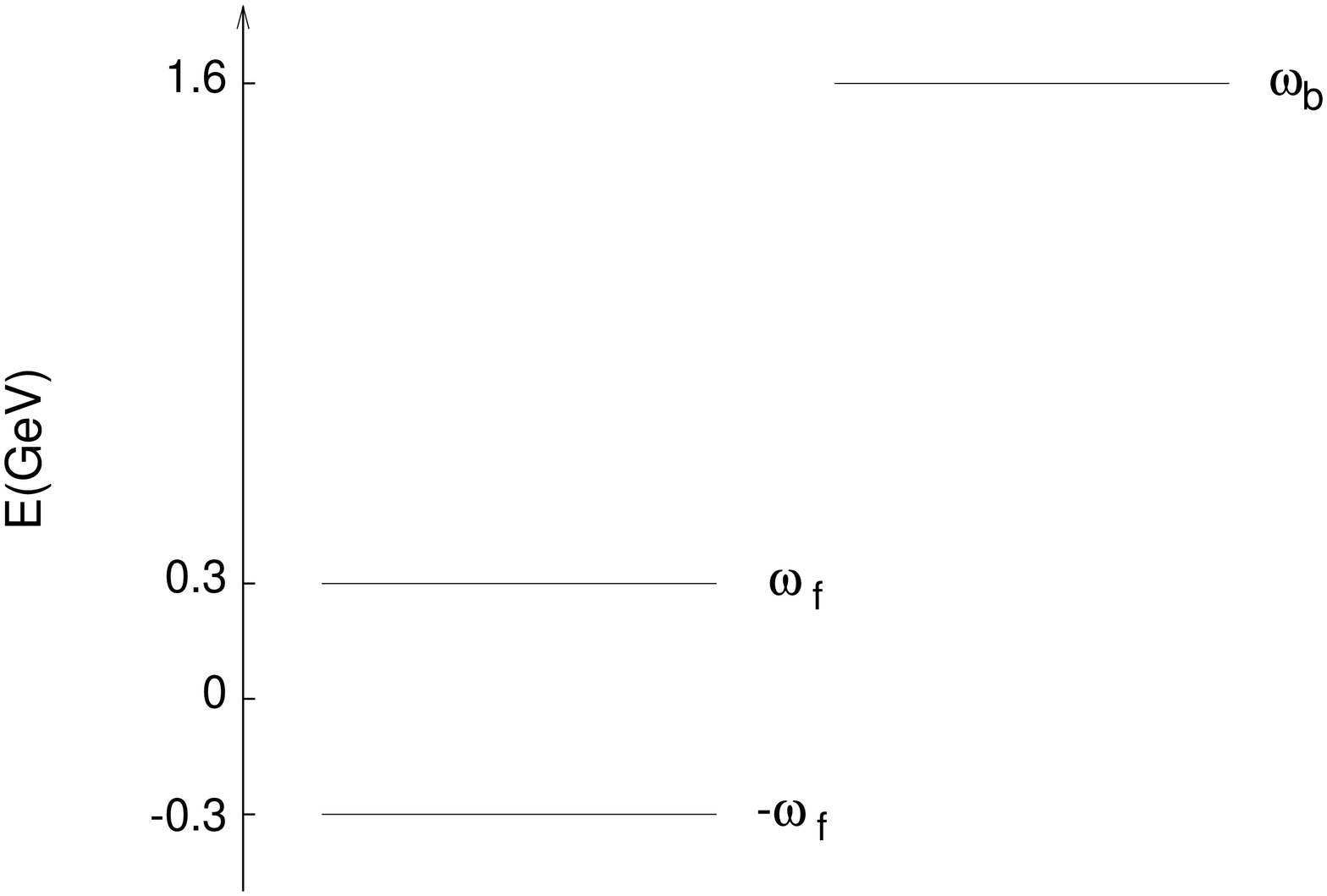}}}
\vskip 0.5cm
\caption{
Schematic representation of
the model space. The fermion levels are indicated by their
energies $\pm \omega_f$. The gluon-pairs are represented by the
level at the energy $\omega_b$.
}
\label{figure1}
\end{figure}

The degeneracy of each fermion level is 2$\Omega$, where $\Omega$
refers to color, flavor and eventually other degrees of freedom.
If only spin ($n_s$), flavor ($n_f$) and color ($n_c$) degrees of
freedom are considered, the value of $\Omega$ is given by the
product $2 \Omega$ $=$ $n_c n_f n_s$. For flavor (0,0) and spin 0
pairs, only, the model has similarities to the one of Ref.
\cite{schutte}, but with a different interaction. In Ref.
\cite{schutte} the pion condensate in nuclei was the dominant
phenomena. In the present case, nucleons are replaced by quarks
and the pions by gluons. The model has some similarities to Ref.
\cite{pittel}, which is also a Lipkin-type model. There, only
quarks were considered  and the interaction conserves their
number.

For zero temperature and no interactions the lower level is filled
by fermions. The creation (annihilation) operators of these
fermions are $\bd{c}^{\dagger}_{\alpha (1,0)f \sigma i}$
($\bd{c}^{\alpha (1,0)f \sigma i}$), in co- and contra-variant
notation for the indices. The symbol $(1,0)f$ refers to the flavor
part, where $(1,0)$ is the SU(3)-flavor notation and $f$ is a
short hand notation for the hypercharge $Y$, the isospin $T$ and
its third component $T_z$. The index $\sigma$ represents the two
spin components $\pm \frac{1}{2}$, the index $i$ $=$ 1 or 2,
stands for the upper or lower level and the index $\alpha$
represents all remaining degrees of freedom, which are at least 3
because of the color degree of freedom (when only color is taken
into account, instead of $\alpha$ we will use the index $c$).
Lowering and raising the indices of the operators introduces a
phase, which depends on the convention used \cite{draayer}, and a
change of the indices to their conjugate values, i.e., the quantum
numbers $(1,0)YTT_z \sigma$ change to $(0,1)-YT-T_z-\sigma$.

The operators, defined above, contain the relevant degrees of
freedom of QCD, i.e. color, spin and flavor. These basic degrees
of freedom appear at all energies, no matter how the resulting
particles are defined, i.e., either in the perturbative or in the
non-perturbative regime. In the non-perturbative regime one
usually denotes them as {\it constituent} or {\it effective}
particles. This is mainly due to the difference in the spatial
properties, while color, spin and flavor have the same meaning as
in QCD. Here, the quarks and antiquarks are {\it constituent}
particles at {\it low energy} and have little in common (except
for the quantum numbers mentioned) with the ones at high energy.
We shall show that a model which contains these basic degrees of
freedom {\it and} which takes into account the dynamic coupling
with gluons can describe the main characteristics of QCD at low
energy.

The quark and antiquark creation and annihilation operators are
given in terms of the operators $\bd{c}$ and $\bd{c}^{\dagger}$

\begin{eqnarray}
\bd{a}^\dagger_{\alpha f \sigma} & = & \bd{c}^\dagger_{\alpha f \sigma 1}, ~~~
\bd{d}_{\alpha f \sigma}  =  \bd{c}^\dagger_{\alpha f \sigma 2}
\nonumber \\
\bd{a}^{\alpha f \sigma} & = & \bd{c}^{\alpha f \sigma 1}, ~~~
\bd{d}^{\dagger ~\alpha f \sigma}  =  \bd{c}^{\alpha f \sigma 2}
~~~,
\label{c}
\end{eqnarray}
which corresponds to the Dirac picture of particles and
antiparticles: quarks are described by fermions in the upper level
and antiquarks by holes in the lower level.

The gluon sector of the model space is described by bosons which
represent pairs of gluons coupled to spin zero. The energy of a
boson state is fixed at the value $\omega_b=1.6$ GeV
\cite{gluons99}, as mentioned before.

The quark-antiquark pairs of the model are given by

\begin{eqnarray}
\bd{C}_{f_1 \sigma_1 1}^{f_2 \sigma_2 2} & = & \bd{B}_{f_1 \sigma_1}^{\dagger f_2 \sigma_2}
= \sum_\alpha \bd{c}^\dagger_{\alpha f_1 \sigma_1 1} \bd{c}^{\alpha f_2 \sigma_2 2}
= \sum_\alpha \bd{a}^\dagger_{\alpha f_1 \sigma_1} \bd{d}^{\dagger \alpha f_2 \sigma_2}
\nonumber \\
\bd{C}_{f_1 \sigma_1 2}^{f_2 \sigma_2 1} & = & \bd{B}_{f_1 \sigma_1}^{f_2 \sigma_2}
= \sum_\alpha \bd{c}^\dagger_{\alpha f_1 \sigma_1 2} \bd{c}^{\alpha f_2 \sigma_2 1}
= \sum_\alpha \bd{d}_{\alpha f_1 \sigma_1} \bd{a}^{\alpha f_2 \sigma_2}
\nonumber \\
\bd{C}_{f_1 \sigma_1 1}^{f_2 \sigma_2 1} & = &
\sum_\alpha \bd{c}^\dagger_{\alpha f_1 \sigma_1 1} \bd{c}^{\alpha f_2 \sigma_2 1} =
\sum_\alpha
\bd{a}^\dagger_{\alpha f_1 \sigma_1} \bd{a}^{\alpha f_2 \sigma_2} \nonumber \\
\bd{C}_{f_1 \sigma_1 2}^{f_2 \sigma_2 2} & = &
\sum_\alpha \bd{c}^\dagger_{\alpha f_1 \sigma_1 2} \bd{c}^{\alpha f_2 \sigma_2 2} =
\sum_\alpha
\bd{d}_{\alpha f_1 \sigma_1} \bd{d}^{\dagger \alpha f_2 \sigma_2} ~~~.
\label{u12}
\end{eqnarray}

The first two equations describe the creation and annihilation of
quark-antiquark pairs. The pairs can be coupled to definite flavor
$(\lambda ,\lambda )$ $=$ $(0,0)$ or $(1,1)$ and spin $S$ $=$ 0 or
1. We shall write, in  this coupling scheme,
$\bd{B}^\dagger_{(\lambda , \lambda )f, S M}$, where $f$ is the
flavor,  $S$ is the spin and $M$ is the spin-projection. The
operators $\bd{B}_{(\lambda , \lambda )f, S M}$ annihilate the
vacuum $|0>$, which can be taken as the configuration where the
lower state is completely filled and the upper one is empty. Note,
that the vacuum state is not uniquely defined \cite{simple}. All
states, which contain only quarks in the upper level and where the
lower level is completely filled (so that antiquarks are not
activated), regardless of color, as for example the three quark
baryon states, are annihilated by $\bd{B}_{(\lambda , \lambda )f,
S M}$.  This property derives from the fact that the operators
$\bd{B}_{(\lambda , \lambda )f, S M}$ contain an antiquark
annihilation operator which anticommutes with all the quark
creation operators. Therefore, the Hilbert space of the model may
be divided into sectors, each one with a different vacuum state
having a given baryon number. The one with the baryon number zero
is the real particle vacuum.

\subsection{Group Theory of the Fermion Part}

From now on, we restrict to $2\Omega = n_sn_cn_f$ $=$ 18 with
$n_s=2$, $n_c=3$ and $n_f=3$, for the spin, color and flavor
degrees of freedom, respectively. The largest group, whose
generators are $c^\dagger_{c_1 f_1 \sigma_1 i} c^{c_2 f_2 \sigma_2
j}$ ($c_i =1,2,3$, $f_i=1,2,3$, $\sigma_i=1,2$ and $i,j=1,2$), is
the $U(4\Omega )$ group. One possible group chain for the
classification of the states, which include the flavor ($SU_f(3)$)
and the spin ($SU_S(2)$) groups, is given by

\begin{eqnarray}
[1^N] &  [h]=[h_1h_2h_3] & [h^T] \nonumber \\
U(4\Omega )  & \supset  U(\frac{\Omega}{3}) ~~~~~~~~~~~ \otimes & U(12) \nonumber \\
&  ~~~ \cup ~~~ & \cup \nonumber \\
&  (\lambda_C,\mu_C)~SU_C(3) ~~~~  (\lambda_f,\mu_f) & SU_f(3) \otimes SU_S(2) ~S,M ~~~,
\label{group1}
\end{eqnarray}
where the irreducible representation (irrep) of $U(4\Omega )$ is the
completely antisymmetric one and $N$ is the number of particles involved.
The upper index in $[h^T]$ refers to the transposed Young diagram
of $[h]$, where the columns and rows are interchanged \cite{hamermesh}.
Due to the antisymmetric irrep $[1^N]$ of $U(4\Omega)$
the irreps of $U(\Omega /3)$ and $U(12)$ are
complementary and the irrep of $U(\Omega /3)$, which is for $\Omega =9$ the
color group, has maximally three rows \cite{hamermesh}.
In the group chain (\ref{group1}) no
multiplicity labels are indicated. There is a multiplicity $\rho_f$
for $(\lambda_f,\mu_f)$ and $\rho_S$ for the spin $S$.
The color labels $(\lambda_C,\mu_C)$ are related to the $h_i$ via
$\lambda_C = h_1-h_2$ and $\mu_C = h_2-h_3$.
The complete state is given by

\begin{eqnarray}
|N, (\lambda_C ,\mu_C), \rho_f (\lambda_f,\mu_f) Y T T_z, \rho_S S M> ~~~,
\label{state}
\end{eqnarray}
where $Y$ is the hypercharge, $T$ is the isospin and $T_z$ its
third component. For meson-like states, the color quantum numbers
to be considered are $(\lambda_C, \mu_C)=(0,0)$. These states will
be located in an elementary volume of about
$\frac{4\pi}{3}$fm$^3$, corresponding to a sphere of radius 1 fm.

To obtain the values of $h_i$ one has to consider all possible
partitions of $N=h_1+h_2+h_3$, which fixes the color. For
colorless states we have $h_1=h_2=h_3=h$. Each partition of $N$
appears only once. The irrep $[hhh]$ of $U(\frac{\Omega}{3})=U(3)$
($\Omega=9$) fixes the irrep of $U(12)$, as indicated in
(\ref{group1}). For the reduction of the irrep of $U(12)$ we have
written a computer code \cite{sergio}, which is available to the
interested reader. As an example, let us consider the $U(12)$
irrep $[3^60^6]$ and the two $U(4)$ irreps $[9^20^2]$ and
$[9720]$, where the first one contains the state where the lower
level is completely filled and the upper one empty, and the second
irrep is the next highest one which contains flavor $(0,0)$ . The
first is accompanied by flavor $(0,0)_1$ and the second one by
$(0,0)_1$ and $(1,1)_1$, where the subindex denotes the
multiplicity. The spin content of $[9^20^2]$ is given by $0_{55}$,
$1_{45}$, $2_{36}$, $3_{28}$, $4_{21}$, $5_{15}$, $6_{10}$,
$7_{6}$,$8_{3}$ and $9_{1}$. The spin content of $[9720]$ is
$0_{81}$, $1_{171}$, $2_{189}$, $3_{135}$, $4_{90}$, $5_{54}$,
$6_{27}$ and $7_{9}$. The lowest dimensional irrep is $[5^24^2]$
with the spin content $0_{1}$ and $1_{1}$.

\subsection{The Boson Mapping}

The explicit construction of the basis states, Eq. (\ref{state})
and the calculation of the matrix elements, become very involved,
which is in conflict with the idea to develop a simple model. A
way out of it is to use a boson mapping of the pair operators
$\bd{B}^\dagger$ and $\bd{B}$ and work in the boson model space.

The basic ingredients of the model are the pair operators, given in
(\ref{u12}). They can be mapped onto boson operators \cite{klein}

\begin{eqnarray}
\bd{B}_{f_1 \sigma_1}^{\dagger f_2 \sigma_2} & \rightarrow &
\bd{b}_{f_1 \sigma_1}^{\dagger f_2 \sigma_2}
\nonumber \\
\bd{B}_{f_1 \sigma_1}^{f_2 \sigma_2} & \rightarrow &
\bd{b}_{f_1 \sigma_1}^{f_2 \sigma_2}
~~~.
\label{bosons}
\end{eqnarray}
where the operators on the right satisfy the normal boson commutation
relations.

\begin{eqnarray}
\left[ \bd{b}_{f_1 \sigma_1}^{f_2 \sigma_2} ,
\bd{b}_{f_3 \sigma_3}^{\dagger f_4 \sigma_4} \right]
= \delta_{f_3f_2} \delta_{f_4f_1}
\delta_{\sigma_3\sigma_2} \delta_{\sigma_4\sigma_1} ~~~.
\label{commutation}
\end{eqnarray}
The exact boson mapping is quite involved, but it can be obtained
in general \cite{klein,hecht,ring}. For the sake of this work, it
is worth to show that the mapping can be performed, indeed. We
shall: i) work from the beginning in the boson space, ii) define a
Hamiltonian which corresponds to the Hamiltonian acting in the
fermion space, and iii) select a basis. The advantage of working
in the boson space is the simplification in getting the matrix
elements (see below). The price to pay is related to the
appearance of {\it non-physical} states \cite{klein}, as we shall
discuss later on.

In order to choose a basis in the boson space, we profit from the
fact that the basic degrees of freedom are given by the boson
creation operators $b^\dagger_{(\lambda , \lambda ) f S M}$ $=$
$b^\dagger_{\lambda f S M}$, with $\lambda$ $=$ 0 or 1 and $S$ $=$
0 or 1. This gives four possible combinations of $[\lambda , S]$:
[0,0], [0,1], [1,0] and [1,1]. Consequently, the total Hilbert
space is the direct product of a one, three, eight and 24
dimensional harmonic oscillators \cite{mosh-oscillator}.

For each harmonic oscillator we can define a seniority basis

\begin{eqnarray}
{\cal N}_{N_{\lambda S}\nu_{\lambda S}} (\bd{b}^\dagger_{\lambda S}
\cdot \bd{b}^\dagger_{\lambda S})^{\frac{N_{\lambda S}-\nu_{\lambda S}}{2}}
|\nu_{\lambda S}\alpha_{\lambda S} > ~~~,
\label{basis}
\end{eqnarray}
where $N_{\lambda S}$ is the number of bosons of type
$[\lambda ,S]$, $\nu_{\lambda S}$ the corresponding seniority
and ${\cal N}_{N_{\lambda S}\nu_{\lambda S}}$ is a normalization constant.
The seniority is the number of $\bd{b}_{\lambda S}$-bosons
{\it not coupled} into pairs.
The $\alpha_{\lambda S}$ contain all other quantum numbers for a
particular harmonic oscillator. The dot in the factor refers to the
scalar product.

The choice of a seniority basis is particularly useful for the
calculation of the matrix elements of the interaction, which
contains expressions of the form $(\bd{b}^\dagger_{\lambda S}
\cdot \bd{b}^\dagger_{\lambda S})$, $(\bd{b}_{\lambda S} \cdot
\bd{b}_{\lambda S})$ and $(\bd{b}^\dagger_{\lambda S} \cdot
\bd{b}_{\lambda S})$, where the latter is just the number operator
of the bosons of the type $[\lambda ,S]$. The exact structure of
$|\nu_{\lambda S} \alpha_{\lambda S}>$ is not needed, except for
the knowledge of the quantum numbers $\alpha_{\lambda S}$.

For the one dimensional harmonic oscillator ([0,0]) the seniority
can take the values 0 or 1. The state is of the form
$(\bd{b}^\dagger_{00})^{N_{00}}|0>$ $=$
$(\bd{b}^\dagger_{00}\bd{b}^\dagger_{00})^{\frac{N_{00}-\nu_{00}}{2}}$
$(\bd{b}^\dagger_{00})^{\nu_{00}}|0>$. For the three dimensional
harmonic oscillator the seniority is equal to the spin $S_{\lambda
S}$. The explicit expression of the state is given in Ref.
\cite{mosh-oscillator}. The eight dimensional oscillator
contributes to flavor only and it is discussed in Appendix A. The
24 dimensional oscillator can be found in Ref. \cite{gluons99},
where the color part in Ref. \cite{gluons99} has to be interpreted
here as the flavor part. In Ref. \cite{gluons99} only singlet
states are listed, but the procedure to obtain non-singlet flavor
states is outlined.

The parity of each state is given by $P=(-1)^{N}$, where $N$
$=$ $\sum_{\lambda ,S} N_{\lambda S}$ is
the total number of bosons. Each boson stems from a particle-antiparticle
pair, which carries negative parity.

In order to obtain the property
under charge conjugation, one has to apply the charge conjugation
operator $\bd{C}$ to the pair creation operator
$\bd{B}^\dagger_{(\lambda , \lambda ) f, S M}$. The result is (see Appendix B)

\begin{eqnarray}
\bd{C}\bd{B}^\dagger_{\lambda f, S M}\bd{C}^{-1} & = & (-1)^S
\bd{B}^\dagger_{\bar{\lambda} \bar{f}, S M} ~~~, \label{c-bdagger}
\end{eqnarray}
where $\bar{\lambda}=(\lambda, -\mu)$, $\bar{f}$ $=$ $-Y$, $T$,
$-T_z$. From this it is clear that only states with $Y=0$,
$T_z=0$, and $\mu=0$, can have a definite $C$-parity. In Eq.
(\ref{c-bdagger}) we make use of the application of the operator
$\bd{C}$ which interchanges quark and antiquark operators
($a^\dagger \leftrightarrow d^\dagger$) and inverts the magnetic
quantum numbers ($Y_i \rightarrow -Y_i$ and $T_{iz} \rightarrow
-T_{iz}$) of flavor and of color only \footnote{After here, for
the sake of notation, we shall indicate charge conjugate states
with a bar on the index f}.

For products of two pair creation operator we obtain

\begin{eqnarray}
\bd{C}\left[ \bd{B}^\dagger_{\lambda_1, S_1}\otimes
\bd{B}^\dagger_{\lambda_2, S_2} \right]^{\rho (\lambda , \mu)}_{f,
M}\bd{C}^{-1} & = & (-1)^{S_1+S_2-\lambda -\mu +\rho_{max} -\rho}
\left[ \bd{B}^\dagger_{\lambda_1, S_1}\otimes
\bd{B}^\dagger_{\lambda_2, S_2} \right]^{\rho (\mu ,
\lambda)}_{\bf{\bar{f}}, M} ~~~, \label{c-bdagger2}
\end{eqnarray}
where $\rho$ is the multiplicity label of $(\lambda , \mu )$ in
the product $(\lambda_1,\lambda_1)\otimes (\lambda_2,\lambda_2)$.
The symbol $\rho_{max}$ denotes  the maximal value of $\rho$. The
phase convention of Ref. \cite{jutta} was used. The symbol
$\otimes$ denotes the combined product in $SU_f(3)$ and $SU_S(2)$.

In analogy, the action of the charge conjugation on a product of
three pair operators can be obtained:

\begin{eqnarray}
& \bd{C}\left[ \left[ \bd{B}^\dagger_{\lambda_1, S_1}\otimes \bd{B}^\dagger_{\lambda_2, S_2}
\right]^{\rho_{12}(\lambda_{12},\mu_{12}), S_{12}} \otimes
\bd{B}^\dagger_{\lambda_3, S_3}\right]
^{\rho (\lambda , \mu), S}_{f, M}\bd{C}^{-1}  = & \nonumber\\
& (-1)^{S_1+S_2+S_3-\lambda -\mu +\rho_{12,max}
-\rho_{12}+\rho_{max} -\rho} \left[ \left[
\bd{B}^\dagger_{\lambda_1, S_1}\otimes \bd{B}^\dagger_{\lambda_2,
S_2} \right]^{\rho_{12}(\mu_{12},\lambda_{12}), S_{12}} \otimes
\bd{B}^\dagger_{\lambda_3, S_3}\right]^{\rho (\mu , \lambda),
S}_{\bf{\bar{f}}, M} & ~~~, \label{c-bdagger3}
\end{eqnarray}
where $\rho_{12}$ is the multiplicity of $(\lambda_{12},\mu_{12})$
in the product of $(\lambda_1,\lambda_1)\otimes
(\lambda_2,\lambda_2)$, $\rho$ is the multiplicity of the total
irrep in the last coupling of the above equation, and
$\rho_{12,max}$ is the maximal value of $\rho_{12}$.

The procedure outlines here can be used in a recursive way for
more involved coupling schemes. For our purpose it is sufficient
to go up to three pairs, which will be the dominant structure at
low energy.

For the G-parity the additional rotation in the isospin space has
to be applied, which changes $T_{i,z}$ to $-T_{i,z}$
\cite{G-parity}. For a polynomial in the pair operators this gives
an additional phase  $(-1)^{T}$, where $T$ is the total isospin
\cite{G-parity}.

The same phase properties under C- and G-parity transformation
have to be valid for the mapped boson operators $\bd{b}^\dagger_{\lambda f, S M}$.

In a seniority basis, the matrix elements are easily obtained, and
they are written

\begin{eqnarray}
< N_{\lambda S}+2 \nu_{\lambda S} \alpha_{\lambda S}|
(\bd{b}^\dagger_{\lambda S} \cdot \bd{b}^\dagger_{\lambda S})
| N_{\lambda S} \nu_{\lambda S} \alpha_{\lambda S}>
& = &
\sqrt{ (N_{\lambda S} - \nu_{\lambda S} +2)
(N_{\lambda S} + \nu_{\lambda S} +d_{\lambda S})}
\nonumber \\
< N_{\lambda S}-2 \nu_{\lambda S} \alpha_{\lambda S}|
(\bd{b}_{\lambda S} \cdot \bd{b}_{\lambda S})
| N_{\lambda S} \nu_{\lambda S} \alpha_{\lambda S}>
& = &
\sqrt{ (N_{\lambda S} - \nu_{\lambda S})
(N_{\lambda S} + \nu_{\lambda S} +d_{\lambda S}-2)}
\nonumber \\
< N_{\lambda S} \nu_{\lambda S} \alpha_{\lambda S}|
(\bd{b}^\dagger_{\lambda S} \cdot \bd{b}_{\lambda S})
| N_{\lambda S} \nu_{\lambda S} \alpha_{\lambda S}>
& = & N_{\lambda S} ~~~,
\nonumber \\
\label{matrixelements}
\end{eqnarray}
where $d_{\lambda S}$ is 1, 3, 8 or 24 for the case of the one,
three, eight or 24 dimensional harmonic oscillator. As a short
hand notation we will use instead of $(\bd{b}^\dagger_{\lambda S}
\cdot \bd{b}^\dagger_{\lambda S})$ the expression
$(\bd{b}_{\lambda S}^\dagger)^2$, and similarly for the other
products, $(\bd{b}_{\lambda S})^2$ and $\bd{b}_{\lambda S}^\dagger
\bd{b}_{\lambda S}$.

As a Hamiltonian, invariant under rotation, charge conjugation
and G-parity, we propose

\begin{eqnarray}
\bd{H} & = & 2\omega_f \bd{n_f} + \omega_b \bf{n_b} + \nonumber \\
& & \sum_{\lambda S} V_{\lambda S}
\left\{ \left[ (\bd{b}_{\lambda S}^\dagger )^2 +
2\bd{b}_{\lambda S}^\dagger \bd{b}_{\lambda S} + (\bd{b}_{\lambda S})^2 \right]
(1-\frac{\bd{n_f}}{2\Omega})\bd{b} + \right.
\nonumber \\
& & \left. \bd{b}^\dagger (1-\frac{\bd{n_f}}{2\Omega})
 \left[ (\bd{b}_{\lambda S}^\dagger )^2 +
2\bd{b}_{\lambda S}^\dagger \bd{b}_{\lambda S}+ (\bd{b}_{\lambda
S})^2 \right] \right\} ~~~. \label{hamiltonian}
\end{eqnarray}
Due to symmetry arguments, the interaction strength $V_{\lambda
S}$ is the same for the two last lines in Eq. (\ref{hamiltonian}).
The term $(\bd{b}_{\lambda S}^\dagger)^2$ ($(\bd{b}_{\lambda
S})^2$) describes the creation (annihilation) of two
quark-antiquark pairs with the simultaneous creation or
annihilation of a gluon pair. The term $\bd{b}_{\lambda S}^\dagger
\bd{b}_{\lambda S}$, in Eq. (\ref{hamiltonian}), describes the
scattering of a fermion pair with the emission or annihilation of
a gluon pair. All processes can be depicted by a Feynman graph and
all graphs can be obtained from any other one  by an appropriate
interchange of lines. Because the strength $V_{\lambda S}$ should
be, basically, invariant under the exchange of lines, we shall use
the same interaction strength for all channels, as a first
approximation. The terms which appear in Eq. (\ref{hamiltonian})
originate in the normal product of $:(\bd{b}_{fS}^\dagger +
\bd{b}_{fS})^2:$, where the square implies a scalar product. The
factor $(1-\frac{\bd{n_f}}{2\Omega})$ represents a cutoff which
can be traced back to the boson mapping of the fermion pairs with
flavor (0,0) and spin 0. This term simulates the effect of an
exact boson mapping \cite{klein,hecht,ring} and it is responsible
for the disappearance of the interaction when the number of pairs
reaches $2\Omega$. In other words, this cutoff term simulates the
Pauli principle which does not allow more than $2\Omega$ pairs.

The Hamiltonian (\ref{hamiltonian}) is the most simple form we can
think of and it contains only four parameters (the values of
$V_{\lambda S}$). The value of $\omega_f$ is fixed to 0.33 GeV,
which is about $\frac{1}{3}$ of the mass of a nucleon. The most
notorious difficulty, associated to the use of the boson mapping,
lies in the Hilbert space of the boson operators. It is larger
than the Hilbert space of the fermion pairs. In some situations
one can identify the source of the spurious dimensions. If, for
example, only flavor (0,0) and spin 0 pairs are taken into
account, the relevant group structure is $U(4\Omega )$ $\supset$
$U(2\Omega )\otimes U(2)$. The irrep of $U(4\Omega)$ has to be
antisymmetric, which implies that the irreps of $U(2\Omega)$ and
$U(2)$ have to be complementary, i.e. if $U(2)$ is given by a
Young diagram with two rows, the one of $U(2\Omega)$ has to be the
adjoint, which is obtained by interchanging rows and columns
\cite{hamermesh}. The upper limit, up to which no spurious states
appear, is $2\Omega$ because $U(2\Omega )$ allows $2\Omega$ rows
in the Young diagram. This is also the maximum number of pairs
allowed, i.e. for this case no un-physical states occur. If flavor
values (0,0) and (1,1) and spin 0 are used, only, we have
$U(4\Omega )$ $\supset$ $U(\frac{2\Omega}{3}) \otimes U(3)$ and up
to $\frac{2\Omega}{3}$ pairs there is no problem with respect to
the appearance of un-physical states. This implies that states
with explicit flavor will present un-physical states only for
large number of bosons. If flavor (0,0) and spin 0 and 1 are
considered, we have $U(4\Omega )$ $\supset$ $U(\Omega) \otimes
U(4)$, which has as an upper limit the number $\Omega$ up to which
no spurious states appear. Finally, for all pairs, i.e. flavor
(0,0), (1,1) and spin 0, 1 the relevant group chain is $U(4\Omega
)$ $\supset$ $U(\frac{\Omega}{3}) \otimes U(12)$ and the upper
limit is $\frac{\Omega}{3}$. This gives us a hint about the group
sequence where un-physical states may appear. The upper limit up
to which all states are physical is lowered in the sequence where
bosons with more degrees of freedom appear. There are states which
can be described either by one or the other type of bosons or even
by a combination of bosons. For example, when both levels, the
lower and the upper ones, are filled there is only one allowed
state, which is the flavor singlet with spin zero. However, all
types of boson pairs can describe it, e.g. when the number of
bosons coupled to flavor singlet and spin zero is equal to
$2\Omega$. In view of these considerations, and for the one,
three, eight and twenty-four dimensional harmonic oscillator
basis, we have introduced the limits $2\Omega$, $\Omega$,
$\frac{2\Omega}{3}$ and $\frac{\Omega}{3}$, respectively. The
higher non physical states do not play an essential role because,
as shown below, the dominant contribution at low energy comes from
configurations with a small number of quark-antiquark pairs
\cite{papII}. By working with these dimensional cut-off values the
influence of non-physical states is minimized. Also, for each
case, the total number of bosons is restricted to $\leq 2\Omega$.
For a reasonable interaction strength, however, the dominant
contribution comes from a small number of bosons. In such cases,
the number of un-physical states is small and they do not
influence much the result. The dimensional cut-off in the Hilbert
space has the advantage that most un-physical states are excluded.
In principle, one can eliminate the spurious states by applying
another, more involved, procedure. For that one has to reduce the
irrep of $U(12)$ to the flavor and spin groups, as done in the
last section. This gives us the allowed content of flavor and spin
for a given irrep of $U(12)$. The matching condition, i.e. by
comparing for a given number of pairs the spin and flavor content
on the boson side to the one on the fermion side, eliminates
un-physical states. If on the boson and fermion sides, for a given
flavor and spin, the number of states are equal, all states in the
model space are taken into account. This is the case of low lying
basis states. A simple counting procedure can be used for other
situations. If there are, for a given flavor and spin, more states
in the boson space than in the fermion space, one can not decide
easily which combination is allowed. However, one can reduce the
number of states of the model space to the same dimension as the
one of the fermion space. As a rule one can first eliminate the
states which contain most of the bosons with the largest degree of
freedom, i.e. which are of the type [1,1], and in this way
proceed, if necessary, until only states with flavor (0,0) and
spin zero bosons are left. At least, the proposed procedure
eliminates most of the spurious states. The error made can be
absorbed in the parameters of the model, a general practice in
dealing with phenomenological models, because in the end the
correct number of degrees of freedom (dimension of the Hilbert
space) dominates in a successful description of the spectrum. The
idea of the proposed procedure is not new and it was used in
another context by J. Cseh et al. \cite{cseh}.

\section{Application to the meson spectrum}

The Hamiltonian (\ref{hamiltonian}) commutes with the isospin
operator and it does not depend explicitly on the hypercharge $Y$.
As a consequence, all states which belong to the same flavor irrep
are degenerate. In principle, we can add terms proportional to
$\bd{T}^2$, $\bd{Y}$ and $\bd{Y}^2$ in order to lift the
degeneracy. These terms will add new parameters to the four
already present ($\omega_b$ is fixed, as in Ref. \cite{gluons99}).
Also, a flavor mixing term could be added, as suggested by the
$\eta$-$\eta^\prime$ mixing \cite{eta}. In order to simplify the
discussion, we shall first ignore these additional interaction
terms.

In fitting the spectrum of the mesons we will use, as an
experimental input for each multiplet, only the state with $T=0$
and $Y=0$. For an octet all states have the same energy as the
isospin singlet and hypercharge zero state. Because later on we
shall take into account flavor mixing interactions too, the
position of the singlet and octet state are not fixed at the
measured energy values but at the values obtained when the flavor
mixing interactions are switched off \cite{G-parity}. The mixing
angle is introduced for two multiplets: the (1,1) $0^-$ and (0,0)
$0^-$ irrep, containing the pions and the $\eta$, $\eta^\prime$,
and the (1,1) $1^-$, (0,0) $1^-$, containing the $\omega$, $\phi$
and $\rho$ particles. The mixing angles are, respectively,
$-23.7^0$ and $35.3^0$, \cite{G-parity}. For other multiplets one
assumes that the mixing angle is zero, because of missing data,
and because of the smallness of the energy splitting between
members, as compared to the energy splitting within the multiplet
which contains the pions or the $\rho$ mesons. The uncorrected
masses for, e.g., the octet (before flavor mixing) are $m_8=615$
MeV in the first case (see notation of Ref. \cite{G-parity}) and
$m_8=940$ MeV for the second case (see also Table 1).

\begin{center}
\begin{table}[th]
\begin{center}
\begin{tabular}{|l|l|l|l|l|}
\hline
particle & $(\lambda_f,\mu_f)$ & $J^\pi$ & $E_{th}$ (GeV) & $E_{\exp}$ (GeV)\\
\hline
vacuum & (0,0) & $0^+$ & 0.0    & 0. \\
$f_0$(400-1200)  & (0,0) & $0^+$ & 0.656  & 0.600 \\
$f_0$(980)       & (1,1) & $0^+$ & 0.797  & 0.980 \\
$f_1$(1420)      & (0,0) & $1^+$ & 1.445  & 1.420 \\
$f_2$(1270)      & (1,1) & $1^+$ & 1.363  & 1.270 \\
$\eta^\prime$ (958)    & (0,0) & $0^-$ & 0.885  & *0.892 \\
$\eta$(1440)   & (0,0) & $0^-$ & 1.379  & 1.440 \\
$\eta$(541)    & (1,1) & $0^-$ & 0.602  & *0.615 \\
$\eta$(1295)   & (1,1) & $0^-$ & 1.428  & *1.295 \\
$\eta$(1760)   & (1,1) & $0^-$ & 1.671  & 1.760 \\
$\omega$(782)  & (0,0) & $1^-$ & 0.851  & *0.861 \\
$\phi$(1020)   & (1,1) & $1^-$ & 0.943  & *0.940 \\
$\omega$(1420) & (1,1) & $1^-$ & 1.389  & 1.420 \\
$\omega$(1600) & (1,1) & $1^-$ & 1.639  & 1.650 \\
\hline
\end{tabular}
\end{center}
\caption{ States used in the fit. The particles are listed in the
first column, and their transformation properties in flavor and
spin are shown in the second and third columns. Note that, for the
particles in the first (0,0), (1,1) $0^-$ and  (0,0), (1,1) $1^-$
irreps, we are listing the value of the masses without flavor
mixing (they are marked by an asterisk). The experimental data are
taken from \cite{particledata} } \label{table1}
\end{table}
\end{center}

In Table 1 we show the states used in the fit. Their flavor, spin
and parity are indicated together with the experimental values. In
total, to perform the fit, we  have considered thirteen states
with spin zero and one in the four parameter fitting procedure.
All other states are predicted, particularly those with spin 2 and
3.

\begin{figure}[tph]

\rotatebox{270}{\resizebox{356pt}{356pt}{\includegraphics[100,130][612,660]{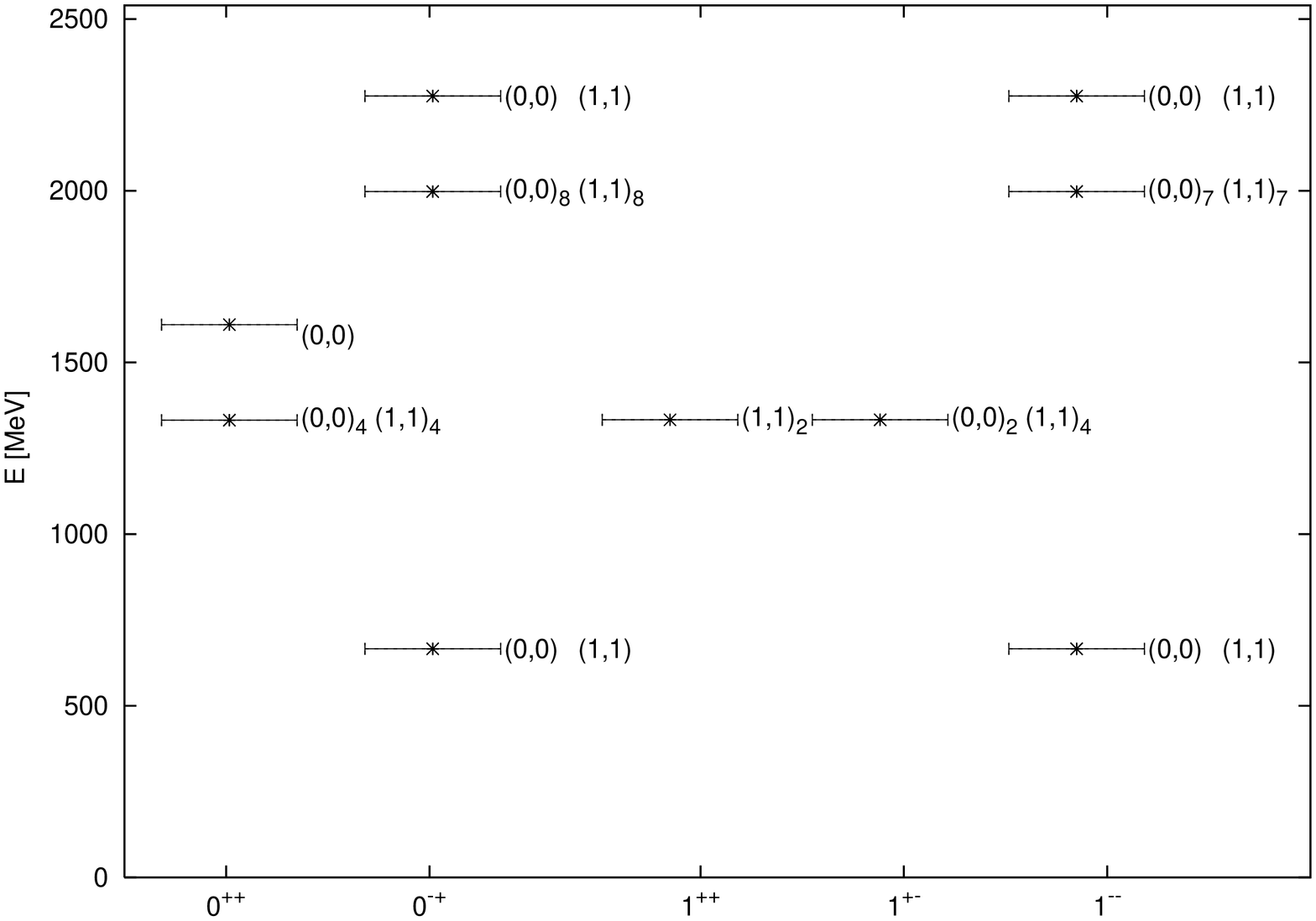}}}
\vskip 0.5cm
\caption{ The meson spectrum for spin 0 and 1 states, for the case
of no interaction. The value $m_f=0.33$ GeV was used. Note the
large multiplicity appearing at already low energies. }
\label{figure2}
\end{figure}

In Figure 2 we give the spectrum for spin 0 and 1 meson states
{\it without} any particle number changing interaction. On each
side of a level the flavor quantum numbers and its degeneracy are
indicated. This serves to illustrate that the multiplicity at
energies lower than 2 GeV is already very large. This is a
consequence of the various manners in which the same set of
quantum numbers can be obtained, for a given configuration, when
many quarks, antiquarks and gluons are considered. This is known
as the {\it multiplicity problem}. The result of the best fit
values, obtained after the interaction is turned on, is given in
Figures 3-6 for spin 0, 1, 2 and 3 respectively. Only states which
correspond to non exotic parity/charge conjugation quantum numbers
are shown. Most of them appear above 2 GeV and some can be deduced
from the gluon spectrum published in Ref. \cite{gluons99}. In
Figures 3-6 each theoretical spectrum is compared to the
experimental one. On the right hand side of each level the
theoretical interpretation in terms of flavor and the multiplicity
is indicated. On the left hand side of each spin ($J^{PC}$) the
experimental information is given, taken from the particle data
group \cite{particledata}. The energy of these states, appearing
in the summary table of Ref. \cite{particledata}, is given in
boxes and the experimental error is reflected by the size of the
box. If the error is very small, the box is replaced by a line.
States which are not in the summary table of \cite{particledata}
are indicated by dashed boxes (lines). Only states which
correspond to isospin singlet and hypercharge zero, after having
corrected for the isospin mixing, are listed.

\begin{figure}[tph]

\rotatebox{270}{\resizebox{356pt}{356pt}{\includegraphics[70,130][612,660]{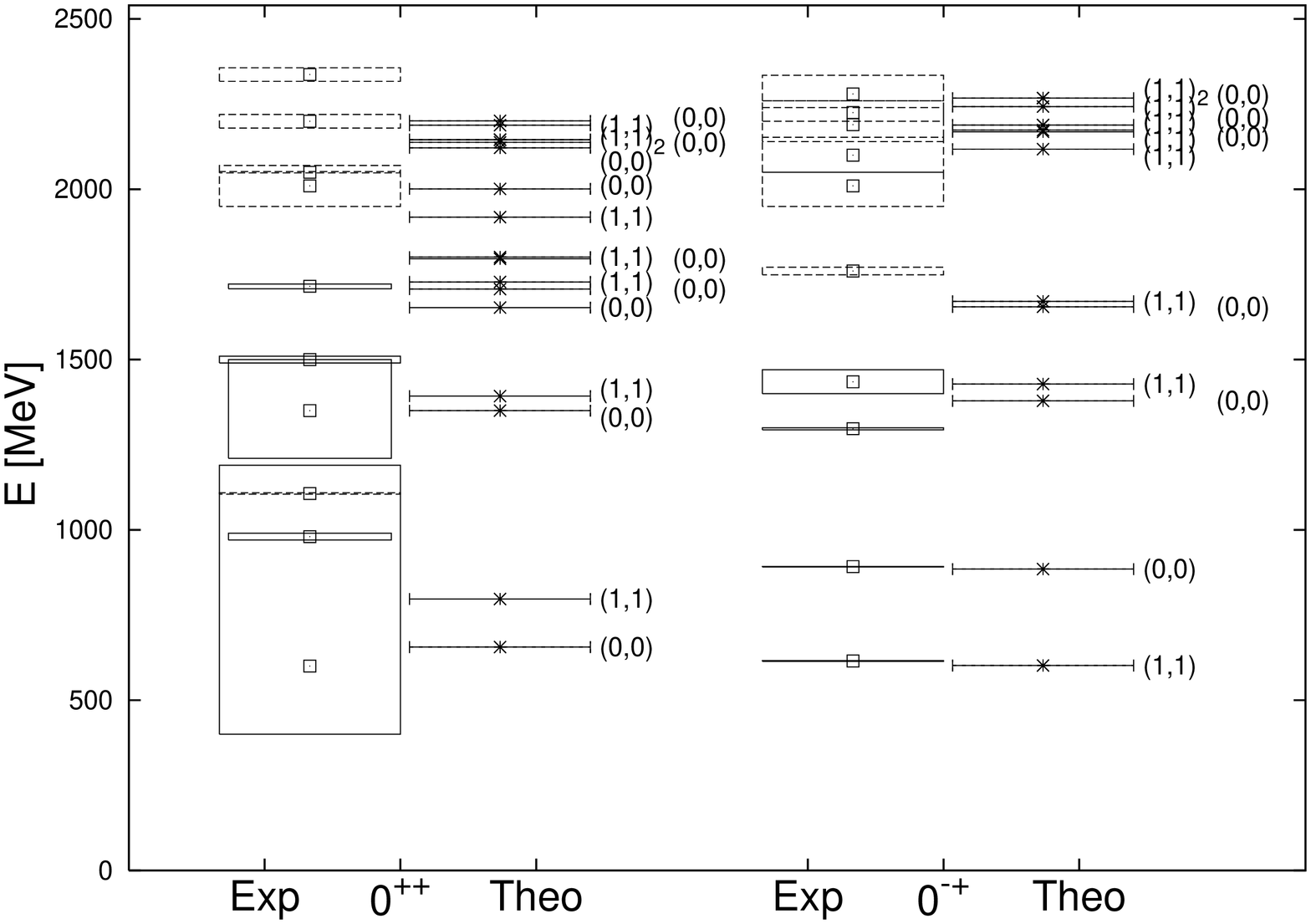}}}
\vskip 0.5cm
\caption{ The meson spectrum for spin 0 as obtained from the fit
to experimental data \cite{particledata}. } \label{figure3}
\end{figure}

\begin{figure}[tph]

\rotatebox{270}{\resizebox{356pt}{356pt}{\includegraphics[100,130][612,660]{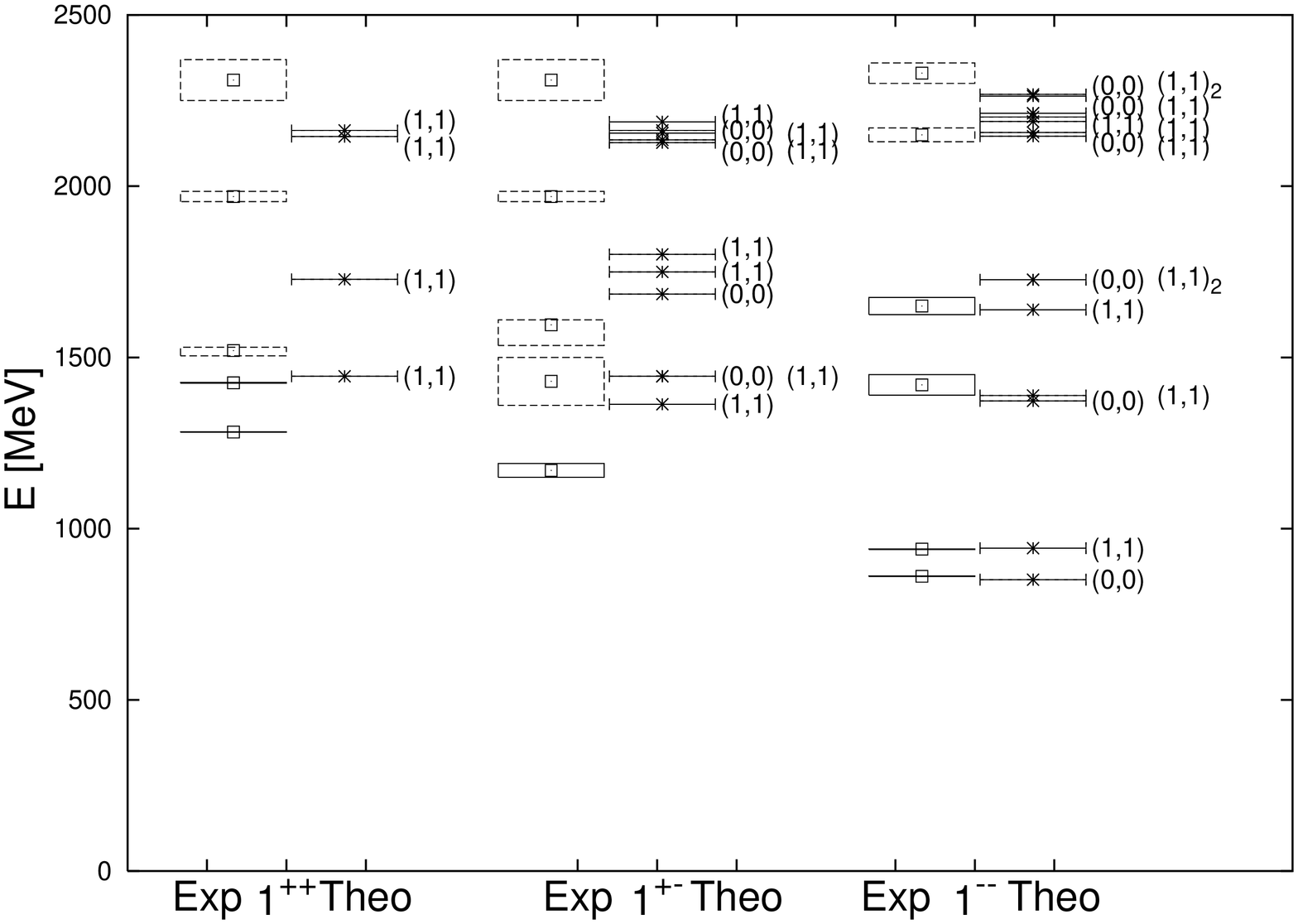}}}
\vskip 0.5cm
\caption{
The meson spectrum for spin 1 as obtained in a fit to experimental data
\cite{particledata}.
}
\label{figure4}
\end{figure}

\begin{figure}[tph]

\rotatebox{270}{\resizebox{356pt}{356pt}{\includegraphics[100,130][612,660]{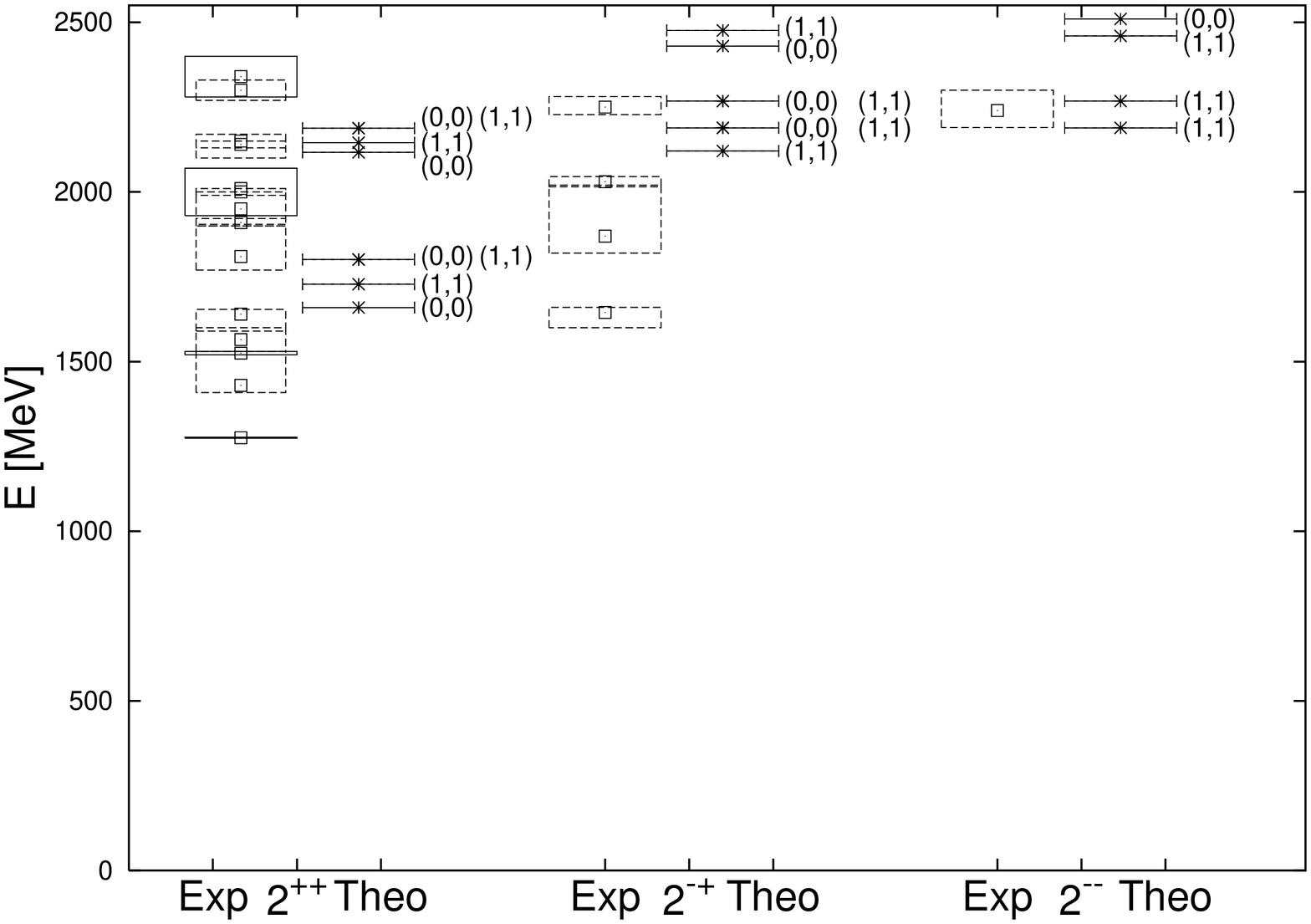}}}
\vskip 0.5cm
\caption{ The meson spectrum for spin 2, obtained with the
paremeters fixed by the fitting procedure. Experimental data are
from \cite{particledata}. } \label{figure5}
\end{figure}

\begin{figure}[tph]

\rotatebox{270}{\resizebox{356pt}{356pt}{\includegraphics[100,130][612,660]{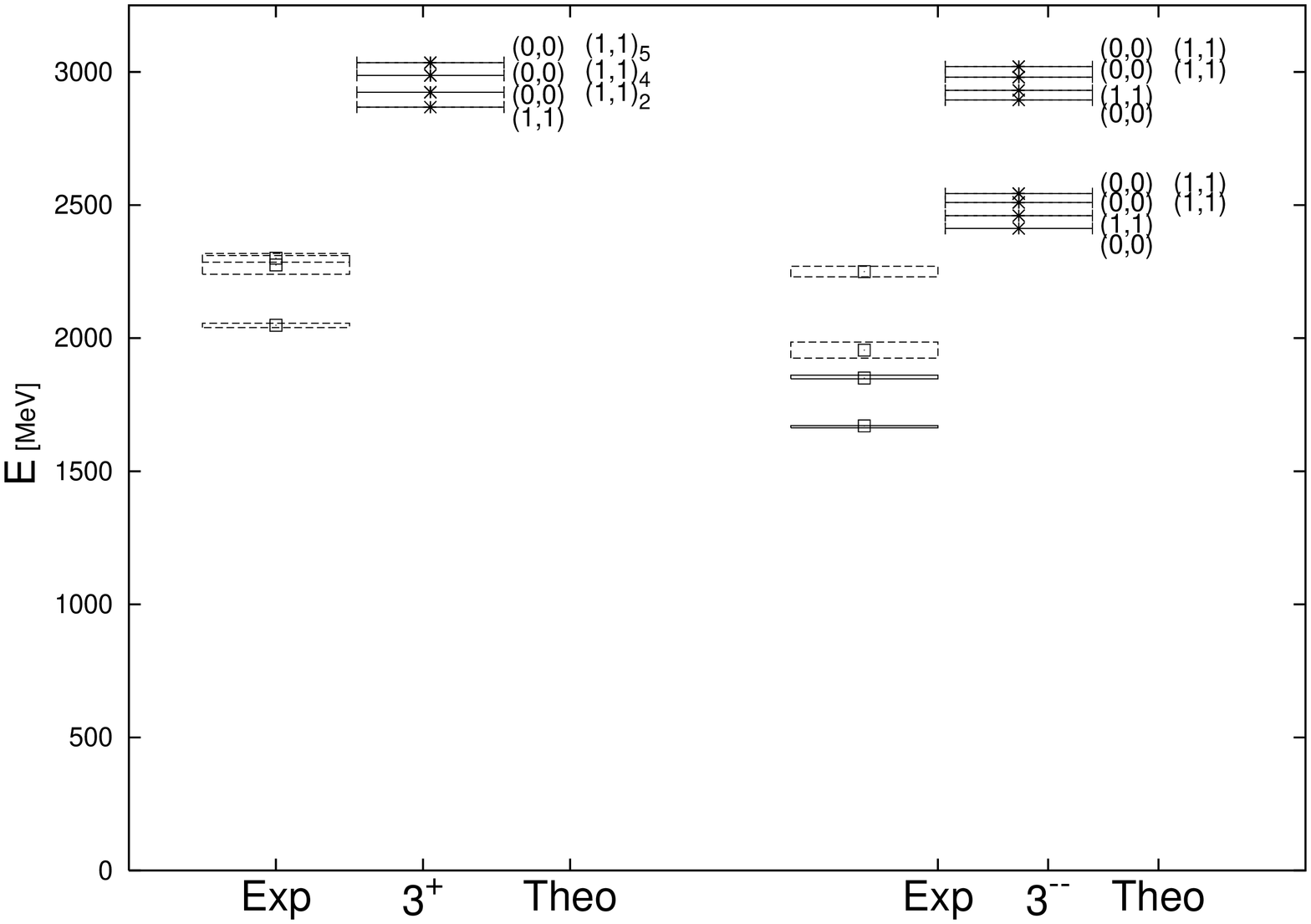}}}
\vskip 0.5cm
\caption{ The meson spectrum for spin 3. Experimental data are
from  \cite{particledata}. } \label{figure6}
\end{figure}

Note that very few states have a multiplicity. Most states were
pushed upwards due to the interaction. This is an effect of the
interaction, because it changes the number of particles and
relates the quark-antiquark sector with the gluons. Models with a
particle conserving interaction will always present the
multiplicity problem. Thus, the particle mixing interaction is
essential to remove the multiplicity problem.

The spin 2 and 3 states where not adjusted. The theoretical
results seems to agree with the data, because the states are
predicted at the correct energy domain. The density of states for
a given flavor also seems to be reproduced.

\begin{center}
\begin{table}[th]
\begin{center}
\begin{tabular}{|l|l|l|l|l|l|l|}
\hline particle & $E_{theo}$ & $(\lambda_f,\mu_f)$ & $J^\pi$ &
$<n_{10}>$ &
$<n_q>$ & $<n_g>$   \\
\hline
vacuum & 0.0    & (0,0) & $0^+$ & 3.118 & 3.177 & 1.705 \\
$f_0$(400-1200)       & 0.656  & (0,0) & $0^+$ & 0.457 & 0.471 & 0.321 \\
$f_0$(980)       & 0.797  & (1,1) & $0^+$ & 3.781 & 3.832 & 1.495 \\
$f_1$(1420)       & 1.445  & (0,0) & $1^+$ & 2.392 & 3.434 & 0.902  \\
$f_2$(1270)       & 1.363  & (1,1) & $1^+$ & 2.464 & 3.519 & 0.993  \\
$\eta^\prime$(958) & 0.885 & (0,0) & $0^-$   & 2.509 & 3.562 & 1.292  \\
$\eta$(1440)            & 1.379   & (0,0) & $0^-$   & 0.773 & 1.790 & 0.444  \\
$\eta$(541) & 0.602       & (1,1) & $0^-$  & 2.711 & 2.766 & 1.163 \\
$\eta$(1295)             & 1.428  & (1,1) & $0^-$   & 1.611 & 1.638 & 0.531  \\
$\eta$(1760)             & 1.671  & (1,1) & $0^-$   & 3.535 & 4.581 & 1.254 \\
$\omega$(782)     & 0.851  & (0,0) & $1^-$   & 2.563 & 3.621 & 1.341 \\
$\phi$(1020)       & 0.943  & (1,1) & $1^-$   & 2.394 & 3.438 & 1.198 \\
$\omega$(1420)     & 1.389  & (1,1) & $1^-$   & 0.853 & 1.870 & 0.468 \\
$\omega$(1600)     & 1.639  & (1,1) & $1^-$   & 3.546 & 4.597 & 1.206 \\
\hline
\end{tabular}
\end{center}
\caption{ Particle content for selected states. In columns we
indicate the theoretical energy ($E_{theo}$), the flavor
($(\lambda_f,\mu_f)$), spin $J$ and parity ($\pi$), expectation
value of the boson pairs in the channel (1,1) $0^-$, ($<n_{10}>$),
expectation value of the total number of quark-antiquark pairs
($<n_q>$) and the total number of gluon pairs ($<n_g>$) with spin
0. } \label{table3}
\end{table}
\end{center}
In Table 2 we show the quark-antiquark pair and gluon pair
contents of some selected states. The total number of
quark-antiquark pairs is denoted by $<n_q>$, where the symbol
$<...>$ indicates the expectation value of this number. The
quantity $<n_{ij}>$ gives the average number of boson pairs of the
type [i,j], while $<n_g>$ is the expectation value of the number
of gluon pairs with spin zero. The total number of gluons is twice
$<n_g>$.

The structure of the calculated ground state (physical vacuum) is
an interesting piece of information about the model predictive
power. The calculated value of the ground state energy is equal to
-0.726 GeV. The physical vacuum state contains about 3.1
quark-antiquark pairs of type [1,0], and the other configurations
contribute with 0.06 pairs. The dominance of the $[1,0]$
quark-antiquark pairs is consistent with the strengths of the
interactions. The parameters, obtained in the fit, are :
$m_f=0.33$ GeV, $V_{0, 0}=0.0337$ GeV, $V_{0, 1}=0.0422$ GeV,
$V_{1, 0}=0.1573$ GeV and $V_{1, 1}=0.0177$ GeV. The channel [1,0]
is clearly the strongest one. Also, the number of gluons is
noticeable, i.e 1.7 pairs, which correspond to more than three
gluons contained in the elementary hadron volume. This indicates a
collective behavior of QCD states at low energy. Indeed, the pion
state (corresponding to $\eta (541)$ in Table 2) contains about
2.7 pairs of the type [1,0] while the rest contributes with about
0.06 pairs, as in the ground state. The number of gluon pairs in
the pion, 1.2 pairs or 2.4 gluons, is similar to the number of
gluon pairs in the ground state. They constitute about 30 per cent
of the particle content. In brief, the calculated states contain a
large number of quark-antiquark pairs and gluons. Roughly
speaking, no single state can be described approximately by a pure
quark-antiquark pair. Note that in this respect theory and
experiment do agree, in spite of the simplicity of the model. We
think that the complex structure of the meson spectrum is
described qualitatively by our model, as well as several other
features, like the position of the first states with spin 2 and 3,
and the density of states with flavor (0,0) or (1,1) for each spin
and parity, charge conjugation number.

Concerning baryons, we have to include them yet in the model. Up
to now they are described as spectators, i.e. without having an
explicit coupling to the quark, antiquark, and gluon sea. For
that, further interaction terms  should be introduced. For
example, one can introduce the interaction

\begin{eqnarray}
\bd{n}_{D,(0,1)0} (\bd{b}^\dagger + \bd{b}) + \bd{n}_{D,(2,0)1}
(\bd{b}^\dagger + \bd{b}) ~~~, \label{barion-int}
\end{eqnarray}
where $\bd{n}_{D,(\lambda ,\mu)~S}$ is the number operator of a
{\it Di-quark} coupled to flavor $(\lambda ,\mu)$ and spin $S$.
This is analogous to the above {\it ansatz} of the Hamiltonian.
The product of two pair creation operators of Di-quarks can not
appear because this would mix the baryon number. The interaction
in (\ref{barion-int}) is a direct extension from the pair operator
interaction of the former Hamiltonian. The first term in
(\ref{barion-int}) acts only on states like the nucleon octet and
the last one on particles like the baryon decouplet. The residual
interaction will mix the number of gluons, and, mainly, increase
the gluon content in the baryons, from a gluon content of about 30
percent to, say, 50 percent. The inclusion of baryons will be
reported in another work \cite{papIII} .

\section{Conclusions}

In the present paper we have advanced a schematic model of QCD,
based on a Lipkin-type model for fermions and an interaction to
one gluon pair states.  We have discussed the low energy structure
of the model. The Hamiltonian is composed by a diagonal, particle
conserving part, and an interaction which couples the
quark-antiquark pairs to the gluons and changes the number of
particles. The model contains only 4 parameters which were
adjusted to reproduce 13 observed meson states with spin  0 and 1.
After fixing these parameters, we have predicted the remaining
part of the spectrum. The complex structure of the meson spectrum
was qualitatively reproduced by our results.

Due to the schematic nature of the model one cannot expect to be
able to reproduce all details of the low energy meson spectrum.
However, the results are in qualitative agreement with data, a
fact that shows the validity of the model as a toy model for QCD.
Baryons where not considered but the extension to this sector was
briefly indicated. The baryons would correspond to states where
three extra quarks are added in the upper level. The corresponding
operators will commute with the boson pair operators and an
interaction of baryon states with the quark-antiquark sea should
be included in the model.

We have found that the inclusion of particle mixing interactions
turns out to be essential in order to remove the multiplicity
problem encountered in other models, when states with many quarks
and antiquarks are considered. This particle changing interaction
also introduces ground state correlations resulting in many
quarks, antiquarks and gluons configurations in the states. It
produces a large contribution of the gluons and the total spin is
not a simple product of a quark-antiquark state but of many
quarks, antiquarks and gluons. This illustrates the fact that,
even at low energy, the structure of the hadron states is by no
means as simple as suggested by earlier particle conservation
models. It also shows that phenomenological potentials, which
simulate the presence of gluons in a pure quark model, cannot
resolve the problem of multiplicity.

\section*{Acknowledgment}
We are pleased to thank Prof. M. Moshinsky for his renewed
interest in our work. We acknowledge financial support through the
CONACyT-CONICET agreement under the project name {\it Algebraic
Methods in Nuclear and Subnuclear Physics} and from CONACyT
project number 32729-E. (S.J.) acknowledges financial support from
the {\it Deutscher Akademischer Austauschdienst} (DAAD) and SRE,
(S.L) acknowledges financial support from DGEP-UNAM. Financial
support from DGAPA, project number IN119002, is acknowledged.

\section*{Appendix A: The eight dimensional oscillator}

The reduction of the eight dimensional oscillator group, $U(8)$,
to the flavor group $SU_f(3)$ is discussed in Ref.
\cite{gluons99}. As an intermediary group, between $U(8)$ and
$SU_f(3)$ one can use the $SO(8)$ group. Though, in Ref.
\cite{gluons99} only the reduction to flavor singlet groups is
listed, the general procedure is outlined. Programs are available
on request \cite{ramon-prog} and the procedure has been published
elsewhere \cite{ramon}.

The generators of the $U(8)$ group are given by
$\bd{b}^\dagger_{(1,1)f_1,00}\bd{b}_{(1,1)f_2,00}$,
where the zeros refer to zero spin and
its projection. Therefore, these bosons can only contribute to the
flavor content. In Table 3 we list the flavor content of up to
four bosons of the type $\bd{b}^\dagger_{(1,1)f,00}$.
As one can see, the multiplicity raises especially for the (1,1) flavor
irrep. With the help of the $SO(8)$ group one can further reduce the
multiplicity. For our purpose this is not necessary.

\begin{center}
\begin{table}
\begin{center}
\begin{tabular}{|c|c|c|}
\hline
$U(8)$ & $SU_f(3)$ & multiplicity \\
$[2]$ & (0,0) & 1 \\
    & (1,1) & 1 \\
    \hline
$[1^2]$ & (0,0) & 0 \\
      & (1,1) & 1 \\
\hline
\hline
$[3]$ & (0,0) & 1 \\
    & (1,1) & 1 \\
    \hline
$[21]$ & (0,0) & 0 \\
      & (1,1) & 3 \\
    \hline
$[1^3]$ & (0,0) & 1 \\
      & (1,1) & 1 \\
\hline
\hline
$[4]$ & (0,0) & 1 \\
    & (1,1) & 2 \\
    \hline
$[31]$ & (0,0) & 0 \\
      & (1,1) & 1 \\
    \hline
$[2^2]$ & (0,0) & 2 \\
      & (1,1) & 2 \\
    \hline
$[21^2]$ & (0,0) & 1 \\
      & (1,1) & 4 \\
    \hline
$[1^4]$ & (0,0) & 1 \\
      & (1,1) & 1 \\
\hline
\end{tabular}
\caption{
The first column gives the Young diagram of the $U(8)$ group,
the second column the irrep of the flavor group $SU_f(3)$
and the third column gives the multiplicity of the flavor irrep.
Only the flavor irreps (0,0) and (1,1) are listed.
}
\end{center}
\end{table}
\end{center}

\section*{Appendix B: Parity, Charge Conjugation and G-Parity}

The charge conjugation operator acts as follows on the quark and antiquark
creation operators

\begin{eqnarray}
\bd{C} \bd{a}^\dagger_{cf\sigma} \bd{C}^{-1} & = & \bd{d}^\dagger_{\bar{c}\bar{f}\sigma}
\nonumber \\
\bd{C} \bd{d}^\dagger_{cf\sigma} \bd{C}^{-1} & = & \bd{a}^\dagger_{\bar{c}\bar{f}\sigma}
~~~,
\label{a-d}
\end{eqnarray}
where the $\bd{a}^\dagger$ transforms in color and flavor as a
(1,0) $SU(3)$ irrep, while the $\bd{d}^\dagger$ transform as
(0,1). If one of the color or flavor index is raised then the
$\bd{d}^\dagger$ transform as (1,0). The "$\bar{c}$" and
"$\bar{f}$" refer to the reflection in the magnetic quantum
numbers of $SU(3)$. I.e. $f$ stands for $Y$, $T$ and $T_z$ and
$\bar{f}$ for $-Y$, $T$ and $-T_z$ and similar for "$\bar{c}$".

With this, the action of the charge conjugation operator on a
quark-antiquark pair is given by

\begin{eqnarray}
&\bd{C} \bd{B}^\dagger_{\lambda f, S M}\bd{C}^{-1}   =
\bd{C}\sum_{cf_1f_2\sigma_1\sigma_2}
\bd{a}^\dagger_{cf_1\sigma_1}\bd{d}^{\dagger~c}_{f_2\sigma_2}
< (1,0)f_1, (0,1) f_2|(\lambda_1,\lambda_1) f>_1 &
\nonumber \\
&  (\frac{1}{2}\sigma_1,
\frac{1}{2}\sigma_2|SM) \bd{C}^{-1} & \nonumber \\
& = \sum_{cf_1f_2\sigma_1\sigma_2}
\bd{d}^\dagger_{\bar{c}\bar{f}_1\sigma_1}\bd{a}^{\dagger~\bar{c}}_{\bar{f}_2\sigma_2}
< (1,0)f_1, (0,1) f_2|(\lambda_1,\lambda_1) f>_1(\frac{1}{2}\sigma_1,
\frac{1}{2}\sigma_2|SM) \nonumber \\
& = -\sum_{cf_1f_2\sigma_1\sigma_2}
\bd{a}^{\dagger~c}_{\bar{f}_2\sigma_2}\bd{d}^\dagger_{c\bar{f}_1\sigma_1}
< (1,0)\bar{f}_1, (0,1) \bar{f}_2|(\lambda_1,\lambda_1) \bar{f}>_1(\frac{1}{2}\sigma_1,
\frac{1}{2}\sigma_2|SM) \nonumber \\
& = -(-1)^{2\lambda_1+1-S}\sum_{cf_1f_2\sigma_1\sigma_2}
\bd{a}^{\dagger~c}_{\bar{f}_2\sigma_2}\bd{d}^\dagger_{c\bar{f}_1\sigma_1}
< (1,0)\bar{f}_2, (0,1) \bd{f}_1|(\lambda_1,\lambda_1) \bar{f}>_1(\frac{1}{2}\sigma_1,
\frac{1}{2}\sigma_2|SM) \nonumber \\
& = (-1)^{S} \bd{B}^\dagger_{\bar{\lambda} \bar{f}, S_1 M} ~~~,
\label{cbdc-1}
\end{eqnarray}
where we made use of the properties of the $SU(2)$ \cite{edmonds} and
$SU(3)$ \cite{jutta} Clebsch-Gordan coefficients.
The subindex 1 in the SU(3) Clebsch-Gordan coefficient indicates a
multiplicity of one \cite{jutta}.

For the product of two pair operators we have

\begin{eqnarray}
& \bd{C}\left[ \bd{B}^\dagger_{\lambda_1, S_1}\otimes \bd{B}^\dagger_{\lambda_2, S_2}
\right]^{\rho (\lambda , \mu)}_{f, M}\bd{C}^{-1}  & \nonumber \\
& = \bd{C}\sum_{f_1f_2M_1M_2} \bd{B}^\dagger_{\lambda_1f_1,S_1 M_1}
\bd{B}^\dagger_{\lambda_2f_2, S_2M_2}
<(\lambda_1,\lambda_1)f_1,(\lambda_2,\lambda_2)f_2|(\lambda,\mu)f>_{\rho} &
\nonumber \\
& (S_1M_1,S_2M_2|SM)\bd{C}^{-1} & \nonumber \\
= & \sum_{f_1f_2M_1M_2} (-1)^{S_1+S_2}\bd{B}^\dagger_{\lambda_1\bar{f}_1,S_1 M_1}
\bd{B}^\dagger_{\lambda_2\bar{f}_2, S_2M_2}
<(\lambda_1,\lambda_1)f_1,(\lambda_2,\lambda_2)f_2|(\lambda,\mu)f>_{\rho} &
\nonumber \\
& (S_1M_1,S_2M_2|SM) & \nonumber \\
= & (-1)^{S_1+S_2-\lambda -\mu +\rho_{max}-\rho}
\sum_{f_1f_2M_1M_2} \bd{B}^\dagger_{\lambda_1\bar{f}_1,S_1 M_1}
\bd{B}^\dagger_{\lambda_2\bar{f}_2, S_2M_2}
<(\lambda_1,\lambda_1)\bar{f}_1,(\lambda_2,\lambda_2)\bar{f}_2|(\lambda,\mu)\bar{f}>_{\rho} &
\nonumber \\
& (S_1M_1,S_2M_2|SM) & \nonumber \\
= & (-1)^{S_1+S_2-\lambda -\mu +\rho_{max} -\rho} \left[
\bd{B}^\dagger_{\lambda_1, S_1}\otimes \bd{B}^\dagger_{\lambda_2,
S_2} \right]^{(\mu , \lambda)}_{\bf{\bar{f}}, M} ~~~, &
\label{cbd2c-1}
\end{eqnarray}
where $\rho$ is the multiplicity index of $(\lambda ,\mu )$ in the product
$(\lambda_1,\lambda_1)\otimes (\lambda_2,\lambda_2)$.

For the product of three pairs we have

\begin{eqnarray}
& \bd{C}\left[ \left[ \bd{B}^\dagger_{\lambda_1, S_1}\otimes \bd{B}^\dagger_{\lambda_2, S_2}
\right]^{(\lambda_{12},\mu_{12}), S_{12}} \otimes
\bd{B}^\dagger_{\lambda_3, S_3}\right]^{(\lambda , \mu), S}_{f, M}\bd{C}^{-1} &
\nonumber \\
& = \sum_{f_{12}f_3M_{12}M_3}
\bd{C} \left[\bd{B}^\dagger_{\lambda_1 S_1}\otimes \bd{B}^\dagger_{\lambda_2S_1}
\right]^{(\lambda_{12},\mu_{12})S_{12}}_{f_{12}M_{12}} \bd{C}^{-1}
\bd{C} \bd{B}^\dagger_{\lambda_3 f_3 S_3M_3} \bd{C}^{-1}  &
\nonumber \\
& <(\lambda_{12},\mu_{12})f_{12},(\lambda_3,\lambda_3)f_3|(\lambda,\mu)f>_{\rho}
(S_{12}M_{12},S_3M_3|SM)\bd{C}^{-1} &
 \nonumber \\
= & \sum_{f_{12}f_3M_{12}M_3}
(-1)^{S_1+S_2-\lambda_{12}-\mu_{12}+\rho_{12,max}-\rho_{12}}
\left[ \bd{B}^\dagger_{\lambda_1S_1}\otimes \bd{B}^\dagger_{\lambda_2S_1}
\right]^{(\mu_{12},\lambda_{12})S_{12}}_{\bar{f}_{12}M_{12}}
(-1)^{S_3} \bd{B}^\dagger_{\lambda_3 \bar{f}_3 S_3M_3} &
\nonumber \\
& <(\lambda_{12},\mu_{12})f_{12},(\lambda_3,\lambda_3)f_3|(\lambda,\mu)f>_{\rho}
(S_{12}M_{12},S_3M_3|SM)\bd{C}^{-1}
& \nonumber \\
= & (-1)^{S_1+S_2+S_3-\lambda_{12}-\mu_{12}+\rho_{12,max}-\rho_{12}}
\sum_{f_{12}f_3M_{12}M_3}
\left[ \bd{B}^\dagger_{\lambda_1S_1}\otimes \bd{B}^\dagger_{\lambda_2S_1}
\right]^{(\mu_{12},\lambda_{12})S_{12}}_{\bar{f}_{12}M_{12}}
\bd{B}^\dagger_{\lambda_3 \bar{f}_3 S_3M_3} &
\nonumber \\
& <(\mu_{12},\lambda_{12})\bar{f}_{12},(\lambda_3,\lambda_3)\bar{f}_3|(\mu,\lambda)\bar{f}>_{\rho}
(-1)^{\lambda_{12}+\mu_{12}-\lambda-\mu+\rho_{max}-\rho}(S_{12}M_{12},S_3M_3|SM)
& \nonumber \\
= & (-1)^{S_1+S_2+S_3-\lambda -\mu
+\rho_{12,max}-\rho_{12}+\rho_{max}-\rho} \left[ \left[
\bd{B}^\dagger_{\lambda_1, S_1}\otimes \bd{B}^\dagger_{\lambda_2,
S_2} \right]^{(\mu_{12},\lambda_{12}), S_{12}} \otimes
\bd{B}^\dagger_{\lambda_3, S_3}\right]^{(\mu , \lambda),
S}_{\bf{\bar{f}}, M} ~~~, & \label{cbd3c-1}
\end{eqnarray}
with the use of the notation of Ref. \cite{jutta} for the $SU(3)$
Clebsch-Gordan coefficients and their symmetry properties.

\end{document}